\documentclass[final,5p,times,twocolumn]{elsarticle} 

\usepackage{graphics}
\usepackage{units}
\usepackage{amssymb}
\usepackage{color}
\usepackage{fixltx2e}
\usepackage{aas_macros}
\usepackage{xfrac}
\usepackage{url}

\makeatletter
\def\ps@pprintTitle{%
 \let\@oddhead\@empty
 \let\@evenhead\@empty
 \let\@evenfoot\@oddfoot}
\makeatother

\begin{document}

\begin{frontmatter}

\title{Design and expected performance of a novel hybrid detector\\for very-high-energy gamma-ray astrophysics}


%
\author[a,b]{P.~Assis}
\author[c]{U. Barres de Almeida}
\author[d]{A.~Blanco}
\author[a,b]{R.~Concei\c{c}\~ao}\corref{cor1}
\ead{\noindent ruben@lip.pt}
\author[e]{B.~D'Ettorre Piazzoli}
\author[f,g,b,a]{A.~De~Angelis}\corref{cor2}
\ead{alessandro.deangelis@pd.infn.it}
\author[h,f]{M.~Doro}
\author[d]{P.~Fonte}
\author[d]{L.~Lopes}
\author[i]{G.~Matthiae}
\author[b,a]{M.~Pimenta}
\author[c]{R.~Shellard}\corref{cor3}
\ead{shellard@cbpf.br}
\author[a,b]{B. Tom\'e}

\cortext[cor1]{Corresponding authors}

\address[a]{LIP Lisboa, Portugal}
\address[b]{IST Lisboa, Portugal}
\address[c]{CBPF, Rio de Janeiro, Brazil}
\address[d]{LIP Coimbra and University of Coimbra, Portugal}
\address[e]{Universit\`a  di Napoli ``Federico II'' and INFN Roma Tor Vergata, Italy}
\address[f]{INFN Padova, Italy}
\address[g]{Universit\`a di Udine, Italy}
\address[h]{Universit\`a di Padova, Italy}
\address[i]{INFN and Universit\`a di Roma Tor Vergata, Roma, Italy}

\begin{abstract}

Current detectors for Very-High-Energy $\gamma$-ray astrophysics
are either pointing instruments with a small field of view (Cherenkov telescopes), 
or large field-of-view instruments with relatively large energy thresholds 
(extensive air shower detectors).

In this article, we propose a new hybrid extensive air shower 
detector sensitive in an energy region starting from about 100 GeV.  The detector combines a small water-Cherenkov detector, able to provide a calorimetric measurement of shower particles at ground, with resistive plate chambers which contribute significantly to the accurate shower geometry reconstruction.

A full simulation of this detector concept shows that it is able to reach better sensitivity than any previous gamma-ray wide field-of-view experiment in the sub-TeV energy region. It is expected to detect with a $5\sigma$ significance a source fainter than the Crab Nebula in one year at $100\,$GeV and, above $1\,$TeV a source as faint as 10\% of it.

As such, this instrument is suited to detect transient phenomena making it a very powerful tool to trigger observations of variable sources and to detect transients coupled to gravitational waves and gamma-ray bursts. 

\end{abstract}


\end{frontmatter}


\section{Introduction}\label{intro}

High energy gamma rays are important  probes of extreme, non thermal, 
events taking place in the universe.
Being neutral, they can cover large distances without being deflected by galactic and 
extragalactic magnetic fields. This feature enables the direct study of their emission 
sources. The gamma emission is also connected to the acceleration of charged cosmic rays and to 
the production of cosmic neutrinos. 
Gamma-rays can also signal the existence of new physics at the fundamental scales, namely by the annihilation or decay of new types of particles, as it is the case for dark matter particles in many models.
This motivation, associated to the advances of technology, has promoted a vigorous 
program of study of high energy gamma rays, with important scientific results 
(see \cite{Pimenta:2015ab,Degrange:2016ab,Funk:2015ab,Hillas:2013am} 
for a summary of the main achievements).

\begin{figure}[ht]
\begin{center}
\includegraphics[width=0.49\textwidth]{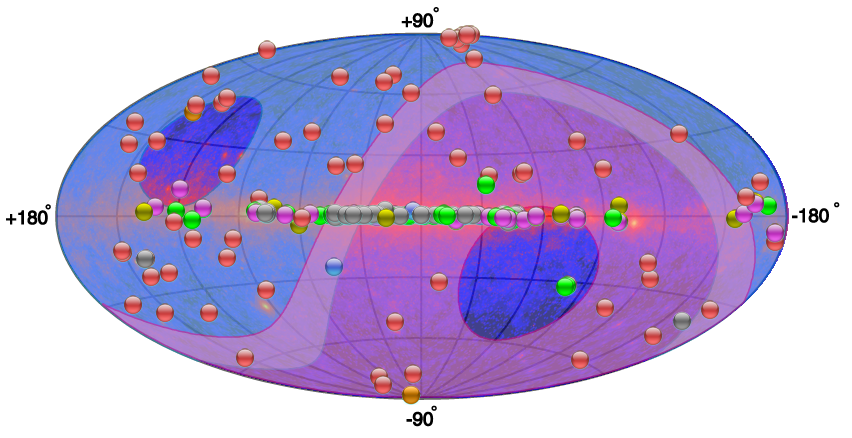}
\end{center}
\caption
{\label{fig:gammamap}Sources of VHE emission displayed in galactic coordinates.
The background represents the high-energy gamma-rays detected by $Fermi$-LAT. 
The clear blue (dominant on the left side) area corresponds to the visible region 
within 30$^\circ$ of the zenith from a detector at a latitude of 22 degrees 
in the Northern hemisphere while the pink (dominant on the right side) shows the corresponding region in the Southern hemisphere.
From http://tevcat.uchicago.edu/, June 2016.}
\end{figure}

The detected sources of cosmic gamma-rays above 30 MeV are concentrated
 around the disk of the Milky Way; in addition there is a set of extragalactic emitters. 
About 3000 sources emitting above 30 MeV were discovered, mostly by 
the Large Area Telescope (LAT) detector \cite{Acero:2015hja}  onboard the $Fermi$ 
satellite, and some 200 of them emit as well above 30 GeV \cite{Ackermann:2016abc} 
(see Fig. \ref{fig:gammamap}) - the region which is labeled the Very High Energy 
(VHE) region.

Our Galaxy hosts about half of the VHE gamma-ray emitters~\cite{Abeysekara:2016ab} and most of them are associated to supernova 
remnants of various classes (shell supernova remnants, pulsar-wind nebulae, etc.). 
The remaining emitters are extragalactic.
The angular resolution of current detectors, which is slightly better than 0.1$^\circ$, does not allow to assign 
the identified extragalactic emitters to any particular region in the host galaxies;  
however, there is some consensus that the signals detected from the Earth must originate in the proximity of 
supermassive black holes at the center of the galaxies \cite{Fuhrmann:2014ab}. 


Still, many problems remain open, of which we may mention:

\begin{itemize}
\item {\it The  origin of cosmic rays} -- supernova remnants (SNRs) are thought to be the sites for the acceleration of protons up to few PeV.
However, the mechanism of acceleration of particles to energies of that order is 
still to be established experimentally. 
The study of the photon yield from Galactic sources for energies larger than 
100~GeV and all the way up to PeV, might solve the problem (see for example 
\cite{Aharonian:2016ab}). 
Actually, photons, which come from $\pi^0$ decay, correspond to hadronic cascades
initiated  at energies at least an order of magnitude larger.

\item {\it The propagation of gamma-rays} -- tells us about their interaction 
with the cosmic background radiation and is a probe to cosmology themes. 

\item {\it New physics} -- the ultimate nature of matter and of physics beyond the 
Standard Model, dark matter or new particles in general, the energy density of the 
vacuum or even quantum gravity may leave imprints in the spectrum of VHE gamma rays.
High-energy gamma-ray astrophysics is sensitive to energy scales important for 
particle physics. For instance, cold dark matter is expected to be found in the 100 GeV scale; supersymmetric particles could appear at the TeV scale; and the Planck scale (an energy $\sim  10^{19}$~GeV, corresponding to a mass $\sqrt{\hbar c/G}$) could be probed indirectly (for discussion see for example  \cite{Pimenta:2015ab}).

\item {\it Transients} -- Many VHE sources are characterised by variability, and it 
is important to be able to detect and measure the corresponding  flares. 
Such flares have a duration that can go from few seconds -- like for the short 
gamma-ray bursts  or the expected counterparts of gravitational waves -- to minutes
for the long gamma-ray bursts, to minutes, hours or even days for the accretion flares of blazars (see e.g. \cite{loeb} and references therein).
\end{itemize}

The layout of this paper is the following.
In the Introduction we have briefly presented the field of very-high-energy 
gamma-ray astrophysics. 
In Section 2 we outline the characteristics of the existing detectors, and we 
explain why a large field-of view detector is needed. 
In Section~\ref{sec:project} we make a case study with a possible design for such 
a detector.
In Section~\ref{sec:data} we describe the characteristics of the gamma-ray signal and of its 
background, and the various Monte Carlo samples used in the analysis of the 
performance of the proposed detector.
In Section~\ref{sec:results} we evaluate the performance of such a new  detector 
using the simulation.
We conclude the paper with final remarks and a summary  in Section~\ref{sec:concl}.


\section{Detection of gamma rays}

The direct detection of primary X/$\gamma$-rays is only possible using satellite-based
detectors since the radiation is absorbed in the atmosphere.
However, the cost of space technology limits the size of satellite-borne detectors to,
roughly, areas of about one square meter.
The largest and most sensitive space-based gamma detector is {\it Fermi}, with 
its high-energy sub-detector, the LAT, with an area of about 3~m$^2$, and an 
effective area -- the product of the geometrical area by the energy dependent detection efficiency -- of 
about 1~m$^2$ for a gamma-ray with an energy of 1 GeV \cite{Atwood:2009abc}.

If the energy of a primary cosmic photon is above some tens of GeV part 
of the products of the air shower, initiated by the interaction of the photon with the atmosphere, can reach the Earth surface.
Ground-based detectors have a large effective area, so their sensitivity is high, 
however they are subjected to a large amount of background events, typically charged 
cosmic rays. They  outperform satellite-based detectors, like the $Fermi$-LAT, in the VHE energy 
region.
The placement of such detectors at high altitude reduces the amount of atmospheric absorption, allowing to reach lower primary energies.

VHE gamma ground based detectors can be divided in two major classes: the Extensive Air 
Shower (EAS) arrays and the Cherenkov telescopes.


\subsection{EAS arrays}

A typical EAS array consists of a large number of detectors that record the secondary shower particles that reach the ground.
Although the energy resolution is poor, it has a very high duty cycle and a large field of view.
Currently, the energy threshold of EAS detectors is at best in the 0.5 TeV - 1 TeV range. At such energies, fluxes are low and large surfaces of the order of~10$^4$~m$^2$ are 
required. This can be achieved:

\begin{itemize}
\item either by using a sparse array of scintillator-based detectors, as for example in 
the  Tibet-AS \cite{Amenomori:2011ab} (for an energy of 100~TeV there are  about 50~000 
electrons in the shower at mountain-top altitudes), which has already finished operations;

\item or by a dense coverage of the ground, to ensure efficient collection and hence 
lower the energy threshold.
The ARGO-YBJ detector \cite{Bartoli:2013qma}\index{ARGO-YBJ detector}, an array of resistive
plate chambers (RPC) at the Tibet site, followed this approach.
\end{itemize}

The largest EAS detector, dedicated to (very) high-energy gamma-rays, presently in operation is the High Altitude Water Cherenkov (HAWC) array 
\cite{Pretz:2015zja, Abeysekara:2015swa} which is in Mexico at an altitude of 4100 m. 

The main problem in this class of detectors is the background rejection, since the flux of 
charged cosmic rays above 0.5 TeV is about a thousand times larger than that from the most 
intense gamma source. Muon detectors are useful instruments to reduce background. 
Background rejection can be otherwise based on the reconstructed shower topology, which is 
different for gamma-initiated atmospheric showers and for hadron-initiated showers.
The direction of the primary particles is estimated from the arrival times with an 
angular precision which can reach about 1~degree at best.

There are new plans to build or expand new, larger and more sensitivity, experiments, for instance, the LHAASO detector \cite{DiSciascio:2016rgi}, with an area of $\sim 90\,000 {\rm m^2}$, which is planned to be built in China. Another EAS gamma-ray experiment extension, worth to be mentioned due to its hybrid detector nature, is the TAIGA experiment~\cite{TAIGA}. 
One interesting aspect is that all the VHE gamma-ray EAS arrays are located at the North hemisphere.



\subsection{Cherenkov telescopes}\index{Cherenkov!telescope}

The Imaging Atmospheric Cherenkov Telescopes \index{Imaging Atmospheric Cherenkov 
Telescope (IACT)}(IACTs) detect the Cherenkov light produced by the atmospheric shower during its development.

The IACTs have currently reached their third generation. They have a low duty cycle (about 1000 to 1500 
hours per year) as they can only operate in moonless nights or with moderate moon light and without clouds. 
These detectors have typically a small field-of-view (FoV) below $5^\circ \times 5^\circ$.
On the other hand, these kind of detectors present a high sensitivity and an energy 
threshold as low as 30 GeV for the present generation.
In fact, most of the experimental results on VHE photons came from IACT observations.



Similarly to EAS arrays, also these experiments have to deal with huge hadronic backgrounds. This is achieved
through the exploration of the shower shape as recorded in the camera~\cite{hillas}.

In order to improve the background rejection, and obtain a better angular and energy resolution, the most recent IACT experiments 
use systems of more than one Cherenkov telescope.

Currently, there are three large operating IACTs: the High Energy Stereoscopic System (H.E.S.S.) 
\index{H.E.S.S. telescope},
the Major Atmospheric Gamma-ray Imaging Cherenkov Telescope (MAGIC),
and the  Very Energetic Radiation Imaging Telescope Array System (VERITAS),
the first one is located in the southern hemisphere, and the two latter in the northern
hemisphere.

The typical sensitivities of these IACTs are such that a source with a luminosity of 1\% of
Crab can be detected at a 5$\sigma$ significance in 50 hours of observation for energies above 50-100 GeV.

The Cherenkov Telescope Array, CTA, is an array of IACTs that will improve the sensitivity to point sources by nearly one order of magnitude to its predecessors. Having a very broad science case, it will be able to survey the whole sky through dedicated scans~\cite{dubus}. This experiment will also be able to extend the current IACT sensitivity up to 100 TeV and a have a broader field-of-view (up to 10 degrees). However, it has a lower duty cycle when compared to EAS-arrays and a still narrower field of view which limits the detection of transient phenomena and the following of extended sources for long term observations.

Table~\ref{tab:DetectorsComparison} compares the main characteristics of currently operating satellite-borne and ground based detectors. 


\begin{table*}
\begin{center}
\begin{tabular}{ | l | l | l | l | }  \hline &&& \\[-.9em]
{\bf Quantity}& {\bf $Fermi$-LAT}    & {\bf IACTs}     & {\bf EAS}    \\ \hline  &&& \\[-.9em]
Energy range & 20 MeV--200 GeV & 100 GeV--50 TeV & 400 GeV--100 TeV \\ &&& \\ [-.9em]
Energy res.  & 5-10\%          & 15-20\%         & $\sim$ 50\%      \\ &&& \\[-.9em]
Duty Cycle   & 80\%            & 15\%            & $>$ 90\%         \\ &&& \\[-.9em]
FoV        & $4 \pi / 5$  & 5 deg $\times$ 5 deg & $4 \pi / 6$      \\ &&& \\[-.9em]
PSF          & 0.1 deg         & 0.07 deg        & 0.5 deg          \\ &&& \\[-.9em]
Sensitivity  & 1\% Crab (1 GeV)& 1\% Crab (0.5 TeV)& 0.5 Crab (5 TeV) \\  \hline
\end{tabular}
\end{center}
\caption{\label{tab:DetectorsComparison}
A comparison of the characteristics of $Fermi$ LAT, of the present IACTs and of a 
typical EAS particle detector array for (very) high-energy gamma-ray detection. 
Sensitivity is computed over one year for $Fermi$ and the EAS, and over 50 hours 
for the IACTs.}
\end{table*}


\section{Design of a novel detector}
\label{sec:project}

Either technique for ground-based observation of gamma-ray sources has clearly 
advantages and drawbacks. In this paper we propose a novel hybrid detector designed as
a ``dense'' array for a surface of 
 $20\,000 {\rm m^2}$,  that can solve some of the limitations of 
the present gamma-ray experiments. 
The requirements of a detector, able to detect possible transients with a large enough sensitivity, would imply a high duty cycle and a large field of view. Moreover, the detector should be able to cover the energy range between 100 GeV and a few TeV.


The detector we are going to propose can be the kernel for an EAS detector in which it is 
complemented by an external sparse detector array covering some 100 000  m$^2$ to increase 
the  sensitivity above 5 TeV and to make it comparable with the foreseen extension of HAWC and 
with the planned LHAASO detector.

ARGO and HAWC have both explored the concept of a single detector composed by: either a
large carpet of RPCs with a very good time and space resolutions (ARGO), or a large
set of Water Cherenkov Detectors WCD, each one with
large volume of water (HAWC).
The ARGO approach relies on a detailed knowledge of the charged particle pattern 
of the air showers at  ground.
The HAWC approach relies on the knowledge of the electromagnetic energy contents 
of the air shower
integrated over a large size region at  ground combined with a good
discrimination power for single muons.

In this paper we argue that a hybrid concept composed by a carpet of low-cost RPCs 
 on  top of WCDs (or other Cherenkov detectors based
on glass or lead glass) of reasonably small dimensions, benefits from
the main advantages of both approaches and can reach a much better sensitivity at the lowest
energies (around 100 GeV). This detector should be placed at high altitude (we assume 5200 m a.s.l. in this paper).

\begin{figure}[ht]
\centering
\includegraphics[width=1.05\linewidth]{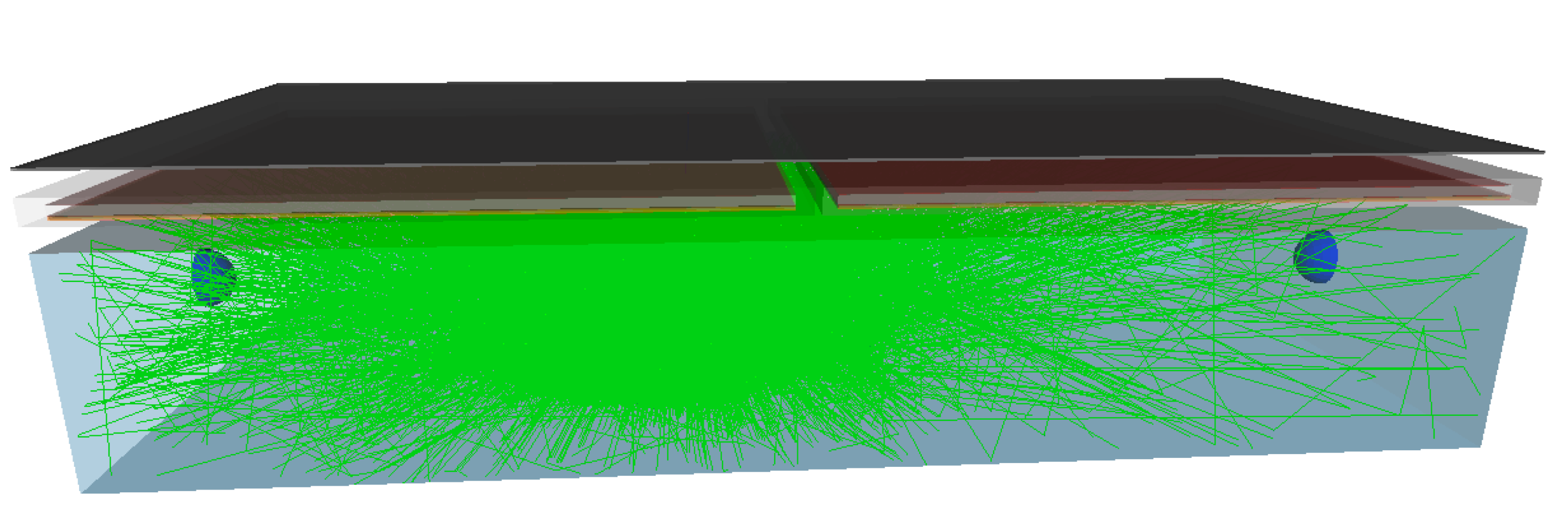}
\caption{
Basic detector station, with one WCD covered with RPCs and a thin slab of lead. 
The green lines show the tracks of the Cherenkov photons produced by the electron and 
positron from the conversion of a photon in the lead slab.
}\label{basic_station}
\end{figure}

Our basic element used in this simulation, the station (Fig. \ref{basic_station}), is 
constructed by putting together one WCD, with a rectangular horizontal surface of 
3~m~$\times$~1.5~m and a depth of 0.5~m, with  signals read by PMTs at both ends of 
the smallest vertical face of the block. The inner walls of the WCD are painted white to maximize the light collection.
On top of the WCD there are two RPCs, each with  a surface of  ($1.5~\times~1.5$)~m$^2$ and 
with 16 charge collecting pads.  
Each RPC is covered with a thin (5.6~mm) layer of lead, to provide the  conversion of shower photons
that have a stronger correlation with the primary direction than the shower electrons.

\begin{figure}
\centering
\includegraphics[width=1.05\linewidth]{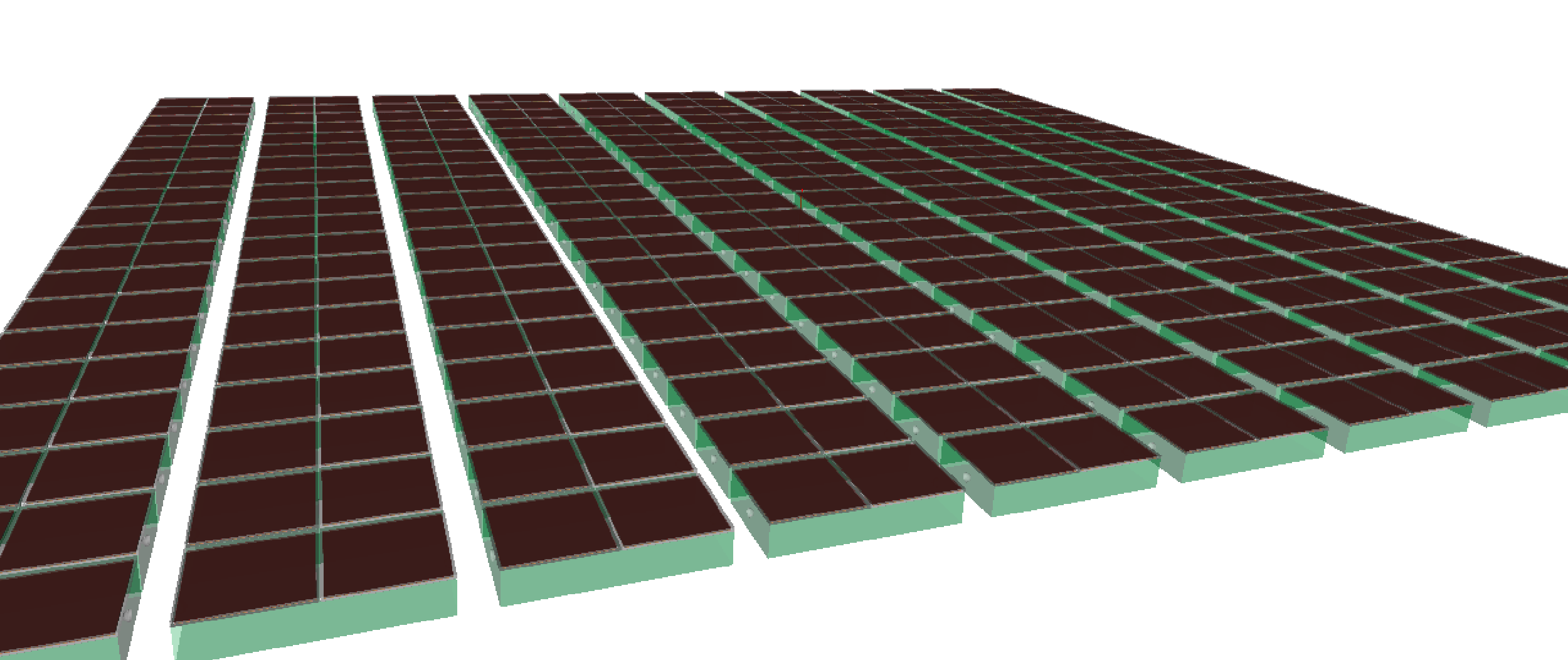}
\caption{
Detail of the layout of the detector used in the case study.
}\label{full_detector}
\end{figure}

The full detector (Fig. \ref{full_detector})  is deployed as an array of individual 
stations set in long lines with each touching the other on their largest dimension. 
The row of lines of detectors are separated by a small distance (roughly 0.5~m) to 
allow access to service the PMTs and the RPCs. 
This arrangement allows for a compact array and for a scaling of the full detector.
The performance results presented herein are based on a baseline configuration 
with an effective area of around 20~000~m$^2$. The station detectors are placed in such a way that the
full array is a circle with radius $\approx 80\,$m.


The proposed RPCs are of the MARTA type (see \cite{Abreu:2015zz}) which have 
been developed in the last four years at LIP in Coimbra, Portugal, and  
successfully tested at Pierre Auger site in Malarg\"{u}e, Argentina. 
These RPCs were designed to work at low gas flux, (1-4)~cc/min, at harsh outdoor 
 environment, and demanding very low maintenance services.
Their intrinsic time resolution  was measured to be better than 1~ns.


\section{Signal, background and simulation tools}\label{sec:data}

We have performed a Monte Carlo simulation of the detector to
evaluate its performance.

For the simulation of atmospheric showers we use the CORSIKA (COsmic
Ray SImulations for KAscade) (version 7.5600) program having the electromagnetic interaction been treated by the EGS4 routines \cite{Heck:1998vt}.
The model to describe hadronic interactions is
FLUKA \cite{FLUKA, FLUKA2}, together with the  QGSJet-II.04 \cite{qgsjetII} model for high-energy
interactions.

Gamma and proton primaries are simulated with a power-law energy spectrum with index -1.0. The energy of the simulated showers ranges between $10\,{\rm GeV}$ and $300\,{\rm TeV}$. Gamma rays are simulated as coming from a point-like source at a zenith angle 
of 10$^\circ$, while protons are simulated with incoming  directions spanning  
the range from 5$^\circ$ to 15$^\circ$ in zenith angle. 

\begin{figure}[ht]
\centering
\includegraphics[width=1.05\linewidth]{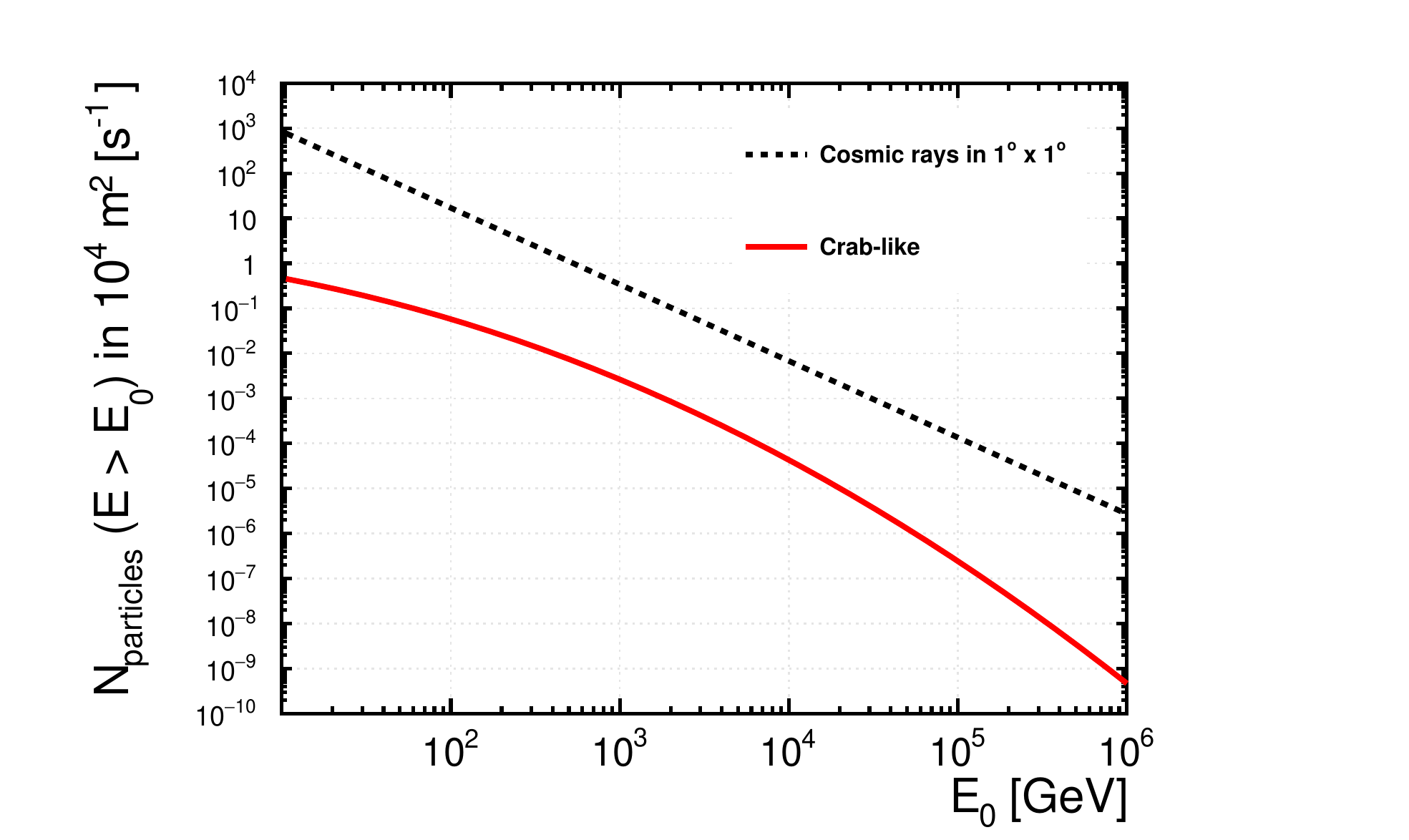}
\includegraphics[width=1.05\linewidth]{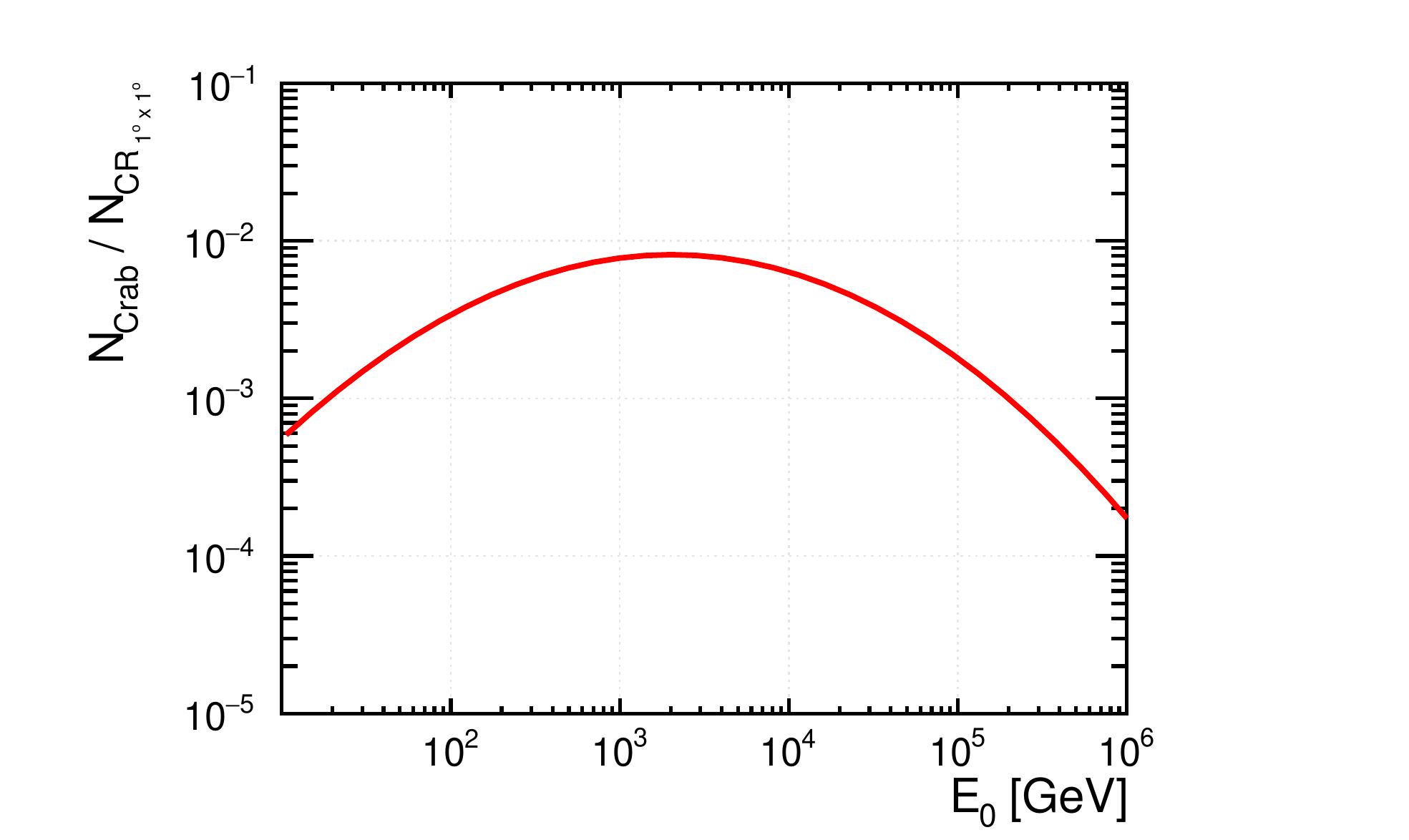}
\caption{Top: Rate of particles above a given energy on a perpendicular surface of 10000 m$^2$, expected 
from a Crab-like source (solid line) and  from the  cosmic-ray proton background in one square degree (dashed line).   
Bottom: ratio signal/background from the above plot.
}\label{fig_sb1}
\end{figure}

Once atmospheric showers have been simulated and the information of
the electrons and photons reaching a height of 5200 m a.s.l. is recorded,
we simulate the response of the detector stations using the GEANT4~\cite{Geant4:2003aa, Geant4:2006aa} toolkit. 

Each simulated shower is reprocessed 100 times,  with a new core position randomly 
set in each realisation. The core sampling area depends on the shower energy to account for the size of it. Showers with energies around $100\,{\rm GeV}$ are uniformly sampled from the center of the array up to $600\,{\rm m}$, while showers of $300\,{\rm TeV}$ are randomised up to $2500\,{\rm m}$.

In the baseline simulation of the WCD units,  the  photomultipliers have a diameter of  15~cm. The maximum quantum efficiency is  30\% at $\lambda \sim 420\, \unit{nm}$.
The inner walls of the tanks are covered with a white diffusive surface in order to increase the light collection. The specular and diffusive properties of this layer are accounted for, as well as its wavelength 
dependence. 
In more detail, the reflectivity was taken to be 95\%, for $\lambda > 450\, \unit{nm}$,  of which  80\% is diffusively reflected  and 20\% is reflected according to a Gaussian angular distribution
 around the specular reflection  direction, with a characteristic width of $\sigma_{\alpha} \sim 0.2^{\circ}$. The detailed structure and materials of the RPC are also described in the simulation.
In particular the information concerning the ionising energy depositions in the gas  is recorded for subsequent processing.   

 The response of the tanks was studied  for single vertical particles injected  uniformly over the top surface. The mean number of photoelectrons for 20~MeV photons,  the median photon energy  at about 10~m from the shower core, is 15 photoelectrons, while for relativistic muons it is 230 photoelectrons. 


In order to evaluate the performance of the detector, we consider a source with an emission  energy distribution like the  Crab Nebula. This nebula ($\sim2$~kpc away) is the brightest steady source detected in VHE gamma-rays \citep{Weekes:1989tc}, and as such is used as a {\it standard candle} in VHE gamma-ray astronomy.
Occasionally the Crab flux can vary in the GeV range as recently observed \cite{Abdo:2011ab, Tavani:2011ab}.

\begin{figure}
\centering
\includegraphics[width=1.15\linewidth]{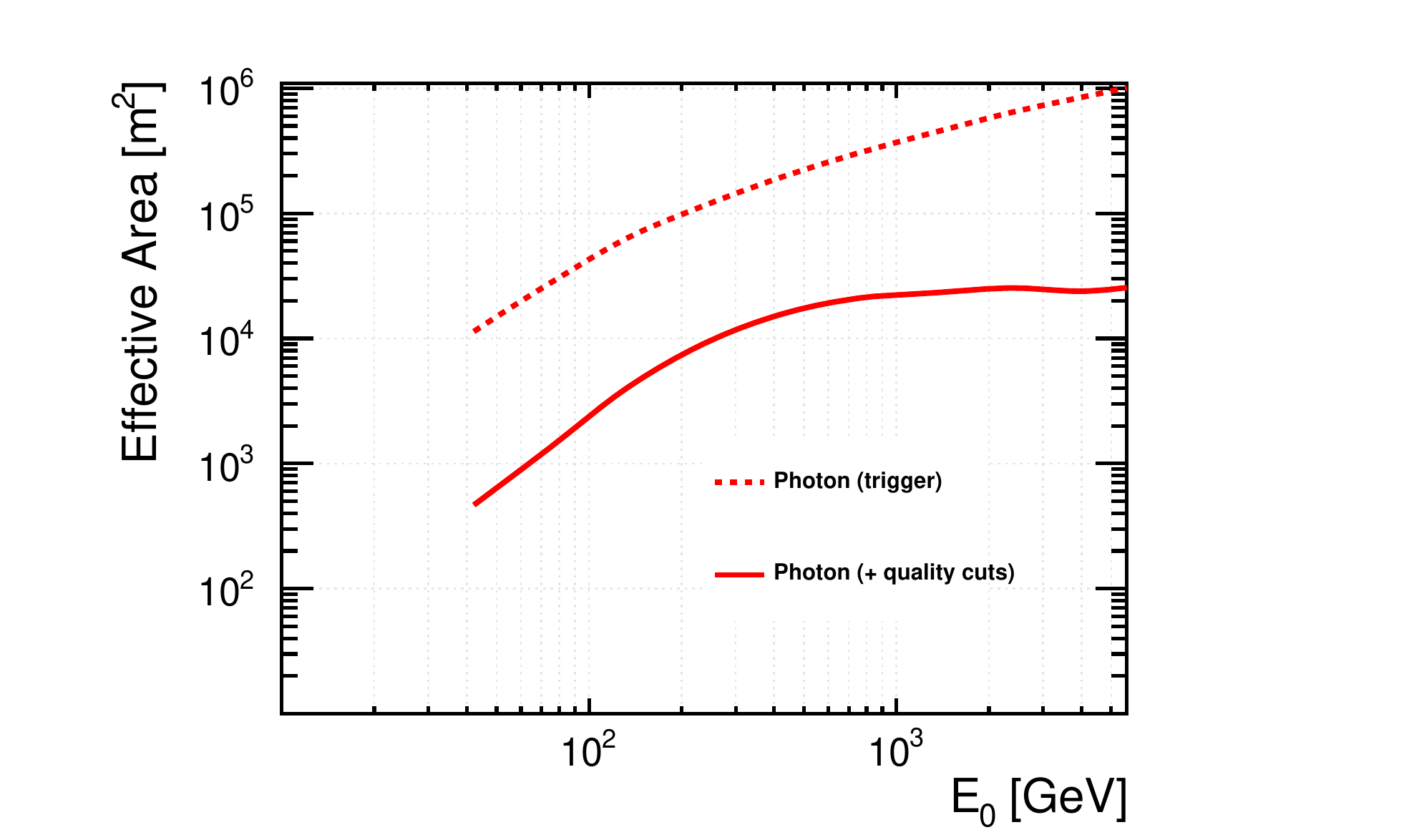}
\caption{Effective area at trigger level (dashed curves) and after the selection used for 
the shower direction reconstruction (solid curves),  for
gamma-ray initiated showers. No g/h separation cuts were applied to any of the curves.
}\label{fig_effarea}
\end{figure}


The {\it stationary}  flux from the Crab Nebula follows, according to
the measurements from MAGIC \citep{Aleksic:2014jva}, a law
\begin{equation}
\frac{dN_\gamma}{dE} \simeq 3.23 \times 10^{-7} \left( \frac{E}{\rm{TeV}}\right)^{  -2.47    -0.24   \left( \frac{E}{\rm{TeV}}\right)}
{\rm{TeV}}^{-1}  {\rm{s}}^{-1}  {\rm{m}}^{-2} \, . \label{eq:ng}
\end{equation}

The  spectral energy distribution of the background cosmic-rays can be obtained from

\begin{equation}
\frac{dN}{dE}  \simeq I_{0}  \,  \left( \frac{E}{\rm{GeV}}\right)^{-2.7}    {\rm{GeV}}^{-1}  {\rm{s}}^{-1} {\rm{sr}}^{-1} {\rm{m}}^{-2}   \, ; \label{eq:nc}
 \end{equation}
 
which is valid from some 10 GeV to a few hundreds of TeV~\cite{PDG}.

The rate of photons from a Crab-like source, above a given energy threshold,
is shown in Fig. \ref{fig_sb1} and compared to the background from cosmic-ray  protons in a square degree.

The photon and proton showers, simulated with an energy spectrum with index -1.0, 
are weighted by $E_{0} \times f(E_{0})$, where  $f$ is the differential 
energy spectrum  in Eqs.~{\ref{eq:ng}} and \ref{eq:nc}, respectively, and  
$E_{0}$ is the energy of the primary particle. 



\section{Estimated performance}\label{sec:results}

\subsection{Effective area}

We use a trigger selection which requires that at least three stations have detected a signal; 
the trigger condition for each station requires at least 5 photoelectrons in each photomultiplier.
Although a detailed study of the trigger is out of the scope of this paper, it is expected that the trigger for such an experiment 
will be composed by a central hardware trigger followed by additional trigger levels which could built up either by hardware or software. 
These trigger levels can be tuned as a function of the intended analysis.  The effective area at trigger level, i.e., the integral of the surface times the trigger efficiency, is shown in Fig. \ref{fig_effarea} for gamma-ray initiated showers. This plot is shown requiring only the trigger and applying the analysis quality cuts that will be described in the following sections. In particular, the reconstructed shower core position is required to be inside the array.
The later results demonstrates the ability to reach an effective area of $\sim 1000\,{\rm m^2}$ for showers with an energy of $100\,{\rm GeV}$.

\subsection{Shower core position reconstruction}

\begin{figure}
\centering
\includegraphics[width=1.05\linewidth]{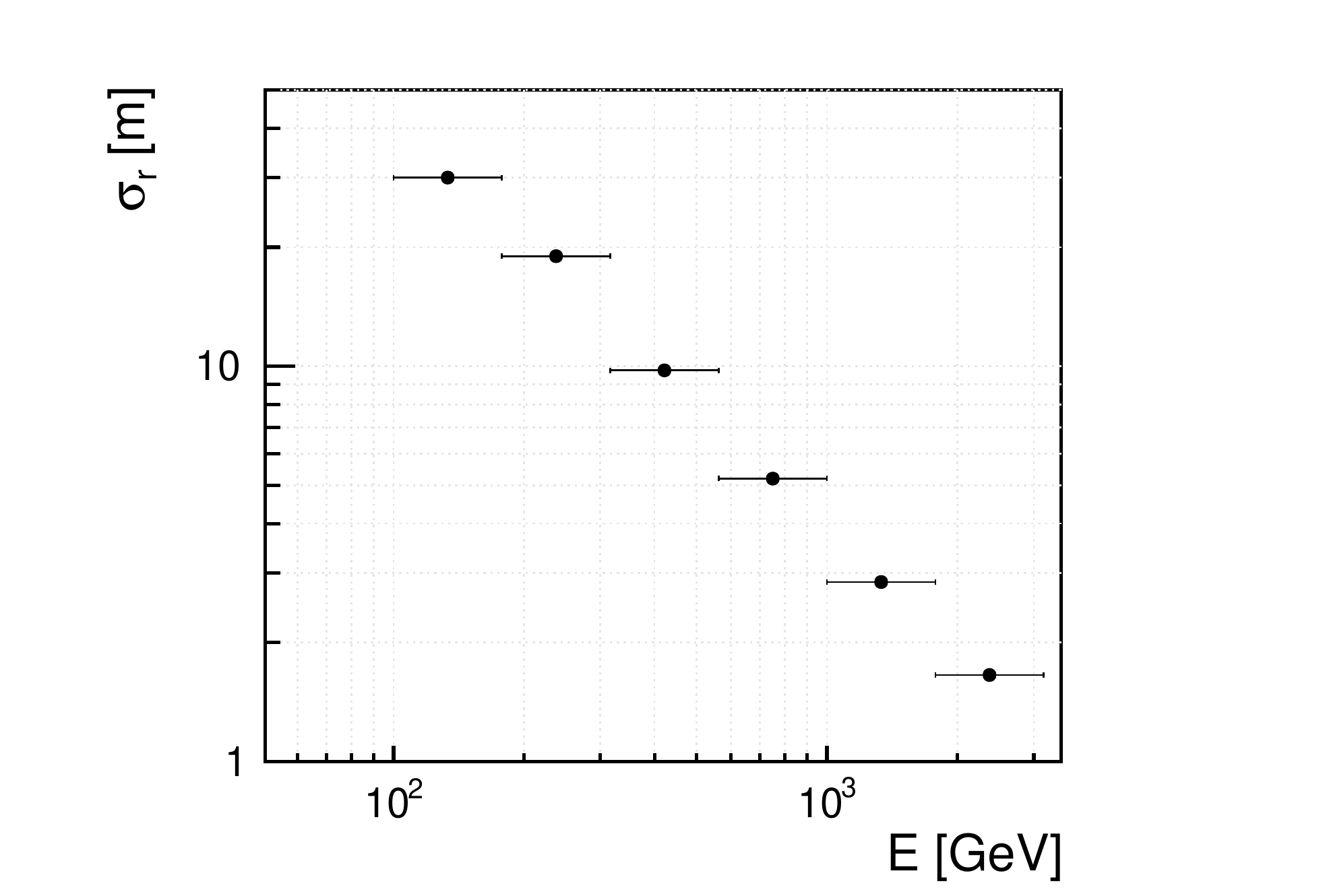}
\caption{Shower core position reconstruction resolution for gamma-ray primaries as a function of the reconstructed energy.}
\label{fig_coreres}
\end{figure}

The shower core position is obtained by using the signal recorded by the WCD stations applying a strategy similar to the one described in~\cite{HAWC2017}. The shower core is obtained by fitting a 2 dimensional lateral density distribution (LDF) to the WCD data. The fitting function parameters were derived using the average gamma 2D-LDF measured at ground by the WCDs. It was found a rather universal behaviour with the shower energy and so the parameters were fixed at all energies. The initial guess to the 2D-fit, which has as free parameters the position ($x_c,y_c$) and a normalisation, is provided by the calculation of the signal barycentre.
The procedure described above allows to obtain a good reconstruction of the core which, as expected, improves as the shower energy increases (see figure~\ref{fig_coreres}). At 200 GeV the core resolution is about $20\,$m, while at $1\,$TeV the shower core can be reconstructed with an accuracy better than $3\,$m.

\subsection{Shower geometric reconstruction}

The arrival direction of the primary particle can be achieved by exploring the arrival time of the shower secondary particles to the ground. For this reason, for the shower geometry reconstruction, the RPCs pad position and hit time will be used. A time resolution of $\sigma_t = 1\,$ns was considered, which can be achieved by present RPCs with standard electronics.
The shower arrival direction is reconstructed assuming a shower front conic model as described in~\cite{ConicFit}. This model has a parameter for the shower curvature which was extracted from gamma shower simulations. Again, no significative evolution with energy was observed and the used parameter is the same for all energies. An iterative fitting process is done until the variation in the reconstructed direction in consecutive iterations is smaller than $0.1$ degree. 
As first estimate for the fit, the direction obtained by assuming a shower front plane model was used.

In order to improve the angular reconstruction it is required that the event has at least 
10 active RPC pads. The pad is only accepted for the reconstruction if it belongs to a triggered WCD station. This cut also reduces the contamination due to low multiplicity accidentals.

Moreover, late arrival hits were removed through the application of a shower front plane model. Hits with a delay bigger than $5\,$ns with respect to the reconstructed shower front are discarded.
\begin{figure}
\centering
\includegraphics[width=1.05\linewidth]{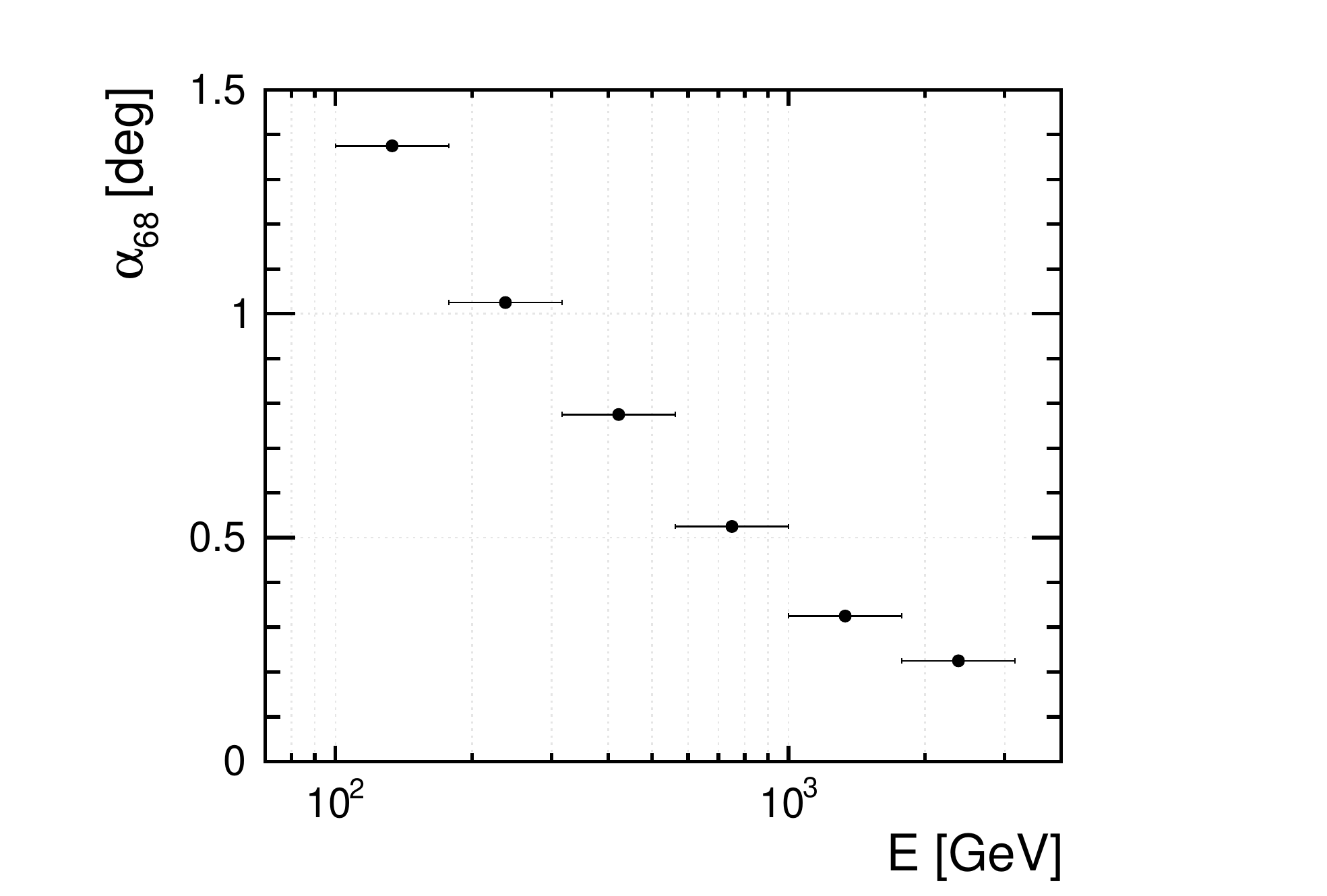}
\caption{Angular resolution for gamma-ray primaries with zenith angle $\theta = 10^\circ$,
as a function of the reconstructed energy.}
\label{fig_angres}
\end{figure}

We compare the reconstructed angle with the simulated angle, and we calculate 
the 68\% containment angle, $\alpha_{68}$. 
The results as a function of the reconstructed energy are shown in Fig. \ref{fig_angres}. 
As expected,  $\alpha_{68}$ decreases with the increase of the reconstructed energy reaching a value of $0.3$ degrees at $E_{rec} = 1\,$TeV. A reasonable resolution, better than $1.5^\circ$, can be achieved at energies around $100$~GeV.



\subsection{Energy estimate}
\begin{figure}[ht]
\centering
\includegraphics[width=1.05\linewidth]{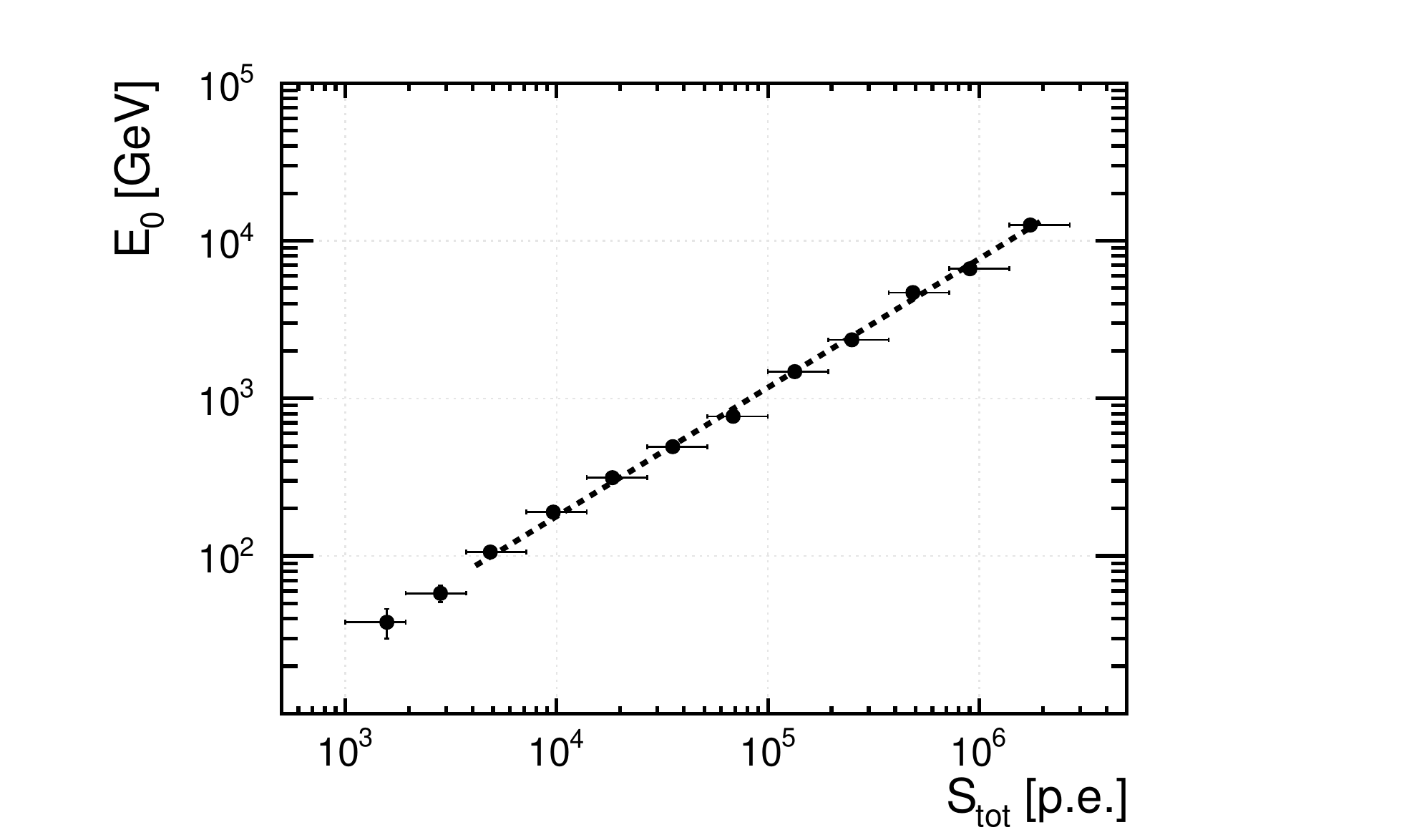}
\includegraphics[width=1.05\linewidth]{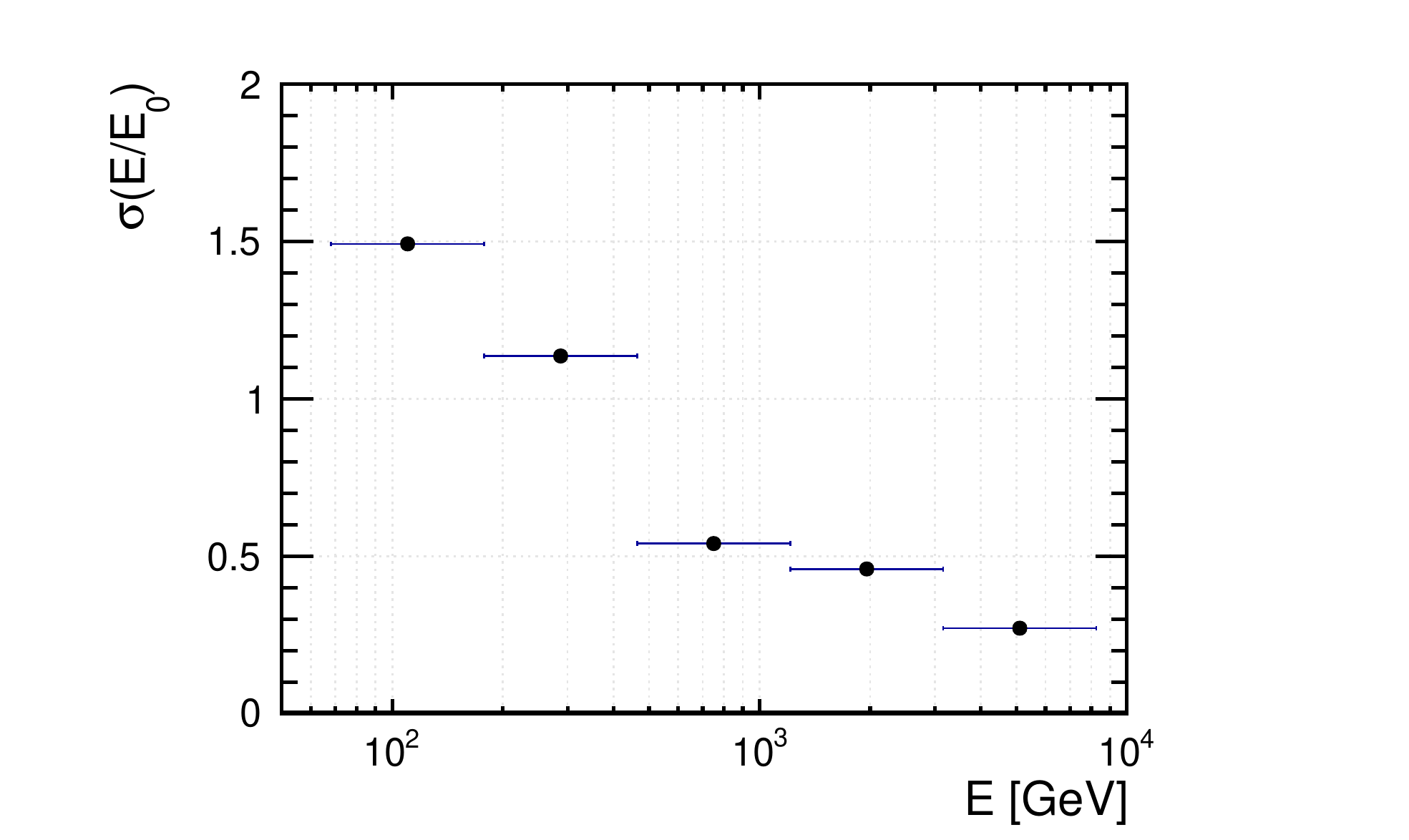}
\caption{Top: Calibration between simulated energy and WCD signal at ground, for photons with a spectral 
energy distribution as for Crab Nebula. 
Bottom: Resolution in the reconstructed energy, for the same sample of photon-initiated 
showers.}
\label{fig_erec}
\end{figure}

The shower energy is reconstructed from the total signal, defined as the sum 
of the number of photoelectrons in all triggered WCD stations. The event is accepted only if the reconstructed core is inside the array.
A calibration curve is obtained using the photon simulation with the Crab 
spectrum, by plotting the median of the generated photon  energies in each bin of  
measured signal, as a function of the median of the measured signal (see Fig.~\ref{fig_erec}, top). 
From this figure it can be seen that a good linearity between simulated energy and total signal can be reached even to energies below $100\,$GeV.

The reconstructed energy follows quite well a
log-normal distribution as a function of the generated energy.
The energy resolution was thus calculated by fitting the distribution of $\ln(E/E_{0})$ 
with a Gaussian function; the relative resolution is shown in Fig. \ref{fig_erec}, bottom. 
The resolution on the reconstructed photon energy depends both on the  detector 
resolution and on the fluctuations in the shower development. 


%
%


\subsection{Hadron background suppression}
The hybrid configuration of the detector units allows to combine the background rejection 
techniques developed by ARGO and HAWC~\cite{Abeysekara:2015swa, Bartoli:2013qma}. 
ARGO gamma/hadron (g/h) discriminators rely on the analysis of patterns at ground. These analyses rely on the use 
of sophisticated tools such as artificial neural networks, which is currently out of the scope of this work. However, 
such analysis could be imported to this detector concept, even if the RPC pads are bigger than the ARGO one's.

Hence, we wish to demonstrate that a good g/h discrimination can be achieved using solely the small water-Cherenkov detector.
Among many observables tested two showed a high potential to distinguish between gamma and hadron showers, which we called: $S_{40}^{high}/S_{40}$ and \emph{compactness}.

The first observable is related to the presence of muons or energetic sub-showers far away from the shower core (above 40 m) in hadronic induced showers, but hardly noticeable in gamma showers. To get a grip on this quantity, we compute the total signal recorded by all WCD stations more than  40 m far from the shower core above a given signal threshold, $S_{40}^{high}$. This quantity, computed event-by-event, is then divided by the total signal present, in the same event, for all WCD stations more than 40 m far from the shower core, $S_{40}$. The signal threshold is taken as the signal that one single muon would give while crossing one WCD. Proton induced showers have in average a higher signal far away from the shower core, which means that the computed ratio ($S_{40}^{high}/S_{40}$ ) will be greater than for gamma primaries. 

The second observable is related to the shower lateral distribution function (LDF) steepness, which is higher for gamma induced showers. An average LDF for gamma 
showers was obtained for each reconstructed energy bin. This LDF, with no free parameters except one normalisation factor, is then fit to the event and the sum of the difference between the WCD data points and the fitted function is used as an estimator for the nature of the primary.
It is worth noting that, while developed independently, similar g/h discrimination strategies were applied by the HAWC collaboration~\cite{HAWC2017}.

In order to maximize the discrimination factor the two discrimination variables were combined using a linear discriminant (Fisher analysis).

\begin{figure}
\centering
\includegraphics[width=0.9\linewidth]{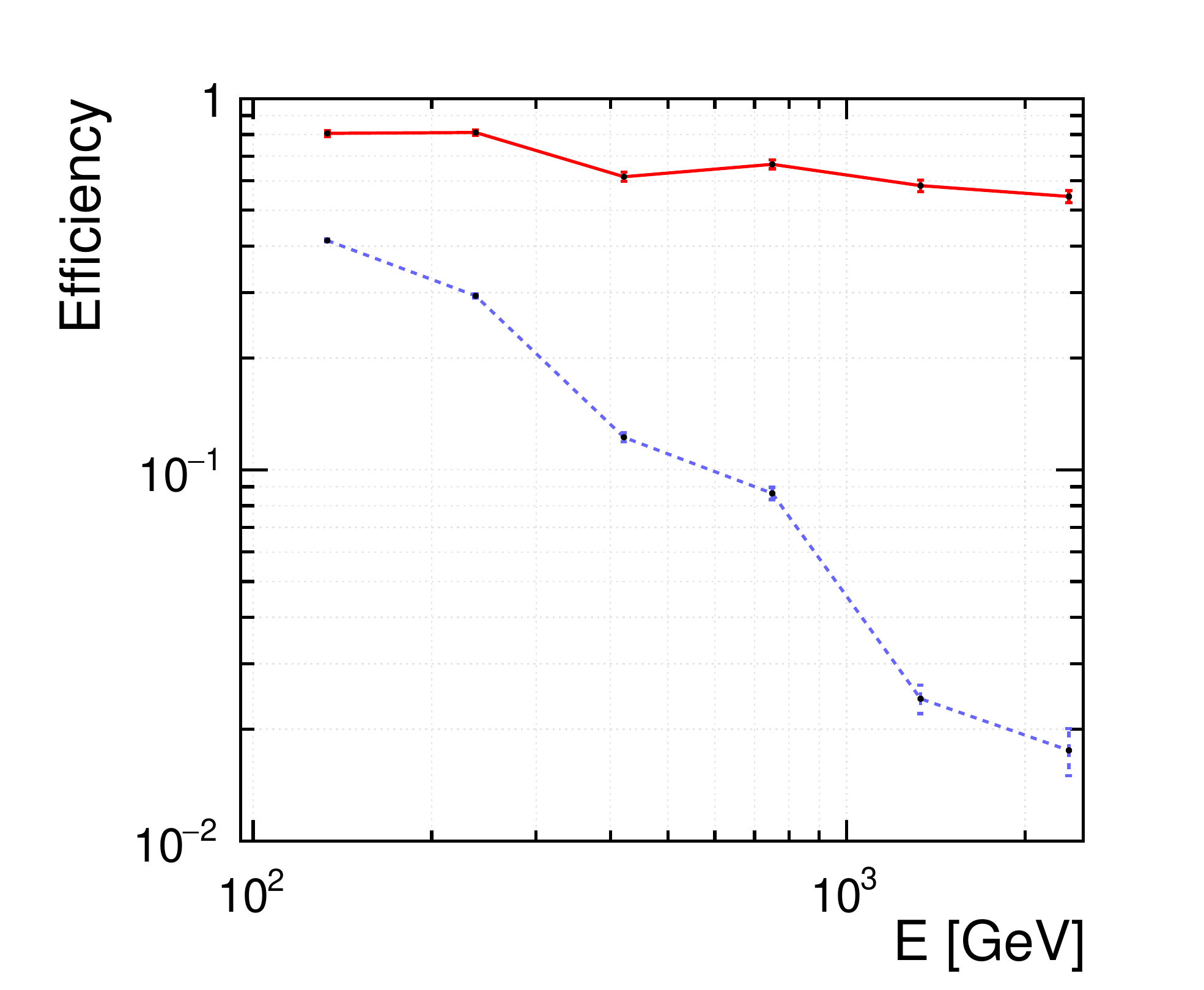}
\caption{Gamma/proton showers selection efficiency as a function of the reconstructed energy. Gammas are shown by the red (full) line while protons appear as blue (dashed).}
\label{fig_ghdisc}
\end{figure}

The results of this exercise are presented in Fig.~\ref{fig_ghdisc}. Here, it is shown the selection efficiency after applying the gamma/hadron cuts. 
For each energy bin, the cut value of the discriminant variable was chosen in order to maximize the  ratio of the signal (gammas) over the square root of the background (protons). From this, it can be seen that the rejection of protons increases rapidly as the shower energy increases, while the gamma selection efficiency remains above 50\%. The calculation of the sensitivity was done using a monotonous smoothed curve of figure~\ref{fig_ghdisc} to reduce statistical fluctuations. The obtained curves for the signal and background efficiencies are within the computed statistical uncertainties.

\subsection{Significance of the Crab signal}

\begin{figure}[ht]
\centering
\includegraphics[width=1.05\linewidth]{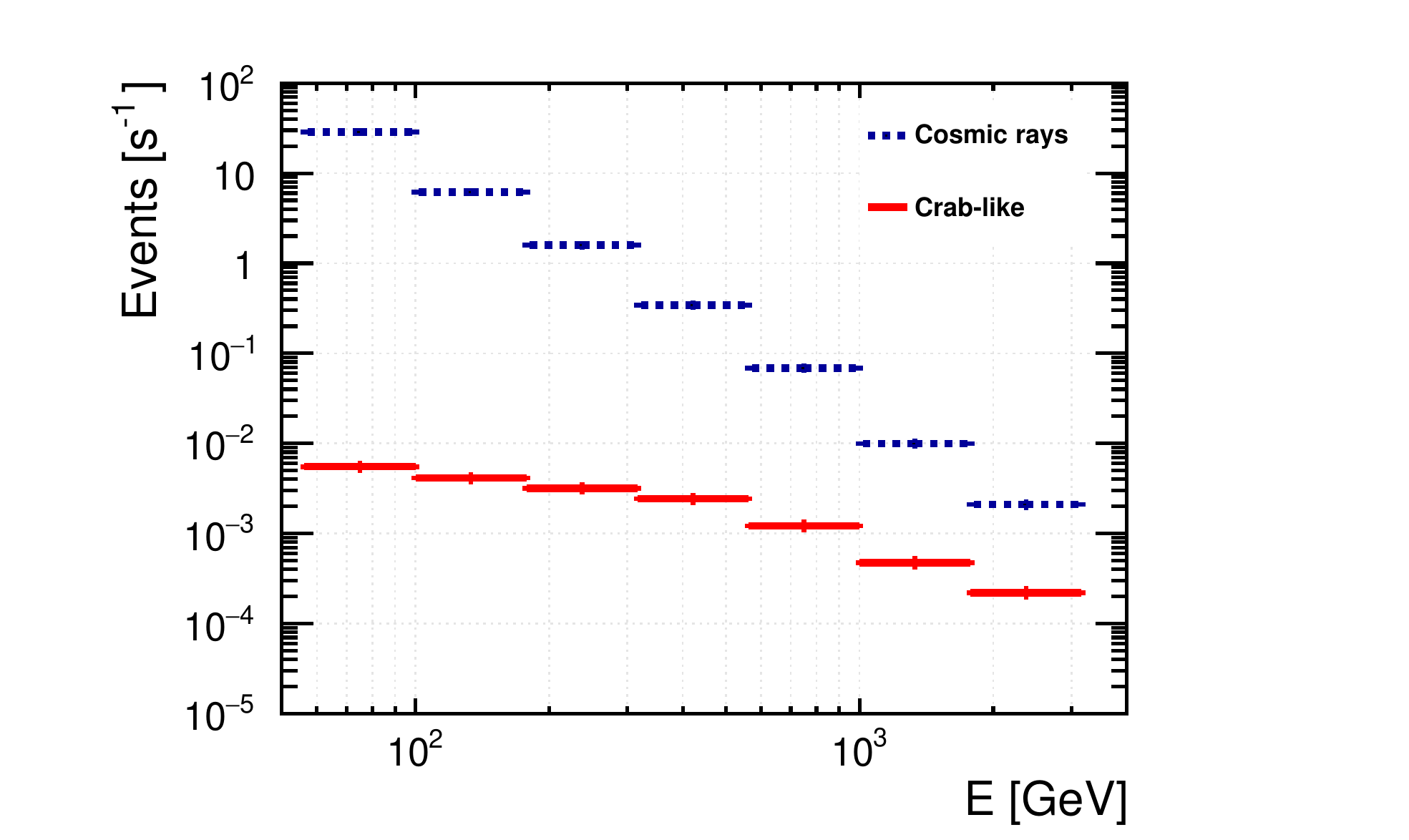}
\includegraphics[width=1.05\linewidth]{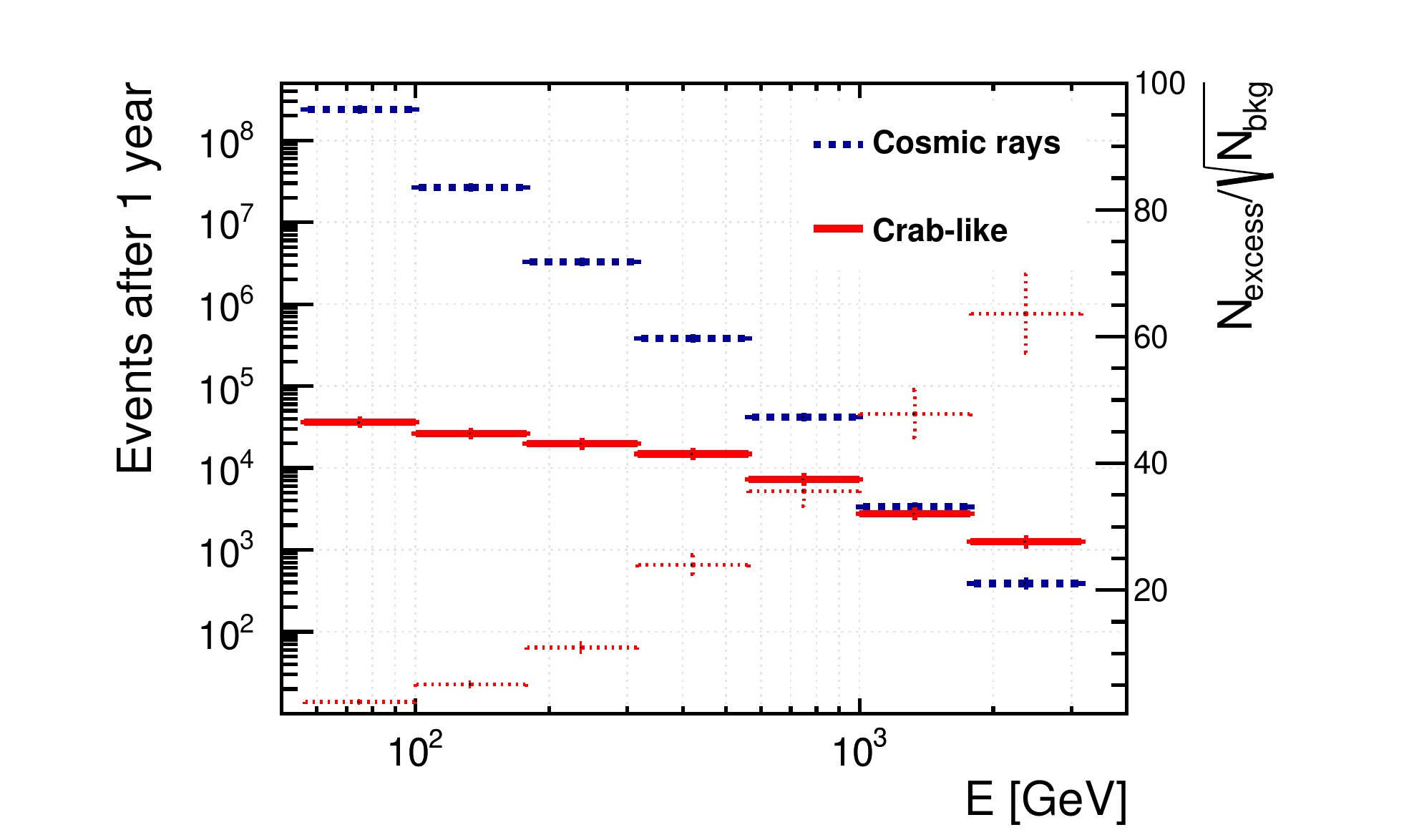}
\caption{
Top: Expected event rate in LATTES  from a Crab-like source  at a zenith angle of 10$^\circ$ (solid line) and  the background from charged cosmic rays 
(dashed line) after the selection in the 1$\sigma$ angular region, 
before the background rejection. 
Bottom: Expected number of events  from a Crab-like source   (solid line) and background (dashed line) in 
one year of effective observation time after all cuts. 
The significance, expressed as the ratio between the signal and the 
square root of the background, is also shown.   
}\label{fig_sb2}
\end{figure}

Gamma-initiated events have been selected within the angular window defined by the cone 
with half-aperture equal to the angular resolution for photons. 
The cosmic-ray background has been calculated for the same window, assuming an isotropic 
flux. Protons are reconstructed with all the analyses procedures developed for gammas. For instance, the calibration curve obtained for photons is used 
to derive the event energy either it is a gamma or a proton.
Due to the limited simulation statistics at the highest energies the sensitivity is only computed up to $2\,$TeV.
The event rate  in each bin of  reconstructed energy, before  background suppression,  
is shown in the  top  plot of  Fig.~\ref{fig_sb2}. 

We then computed the number of events for one year of effective time, after applying the 
hadron suppression efficiency curves; the result is  shown separately for signal and 
background events in the bottom plot of 
Fig.~~\ref{fig_sb2}.   
One year of effective time corresponds to  $7.9 \times 10^{6}$ seconds, assuming 
a duty cycle of 25\% (which corresponds to the average fraction of time at which a source 
culminating at zenith is seen within an angle of 30$^\circ$ from zenith). 

The significance of a detection in terms of number 
of standard deviations $n_\sigma$ can be calculated with a simplified formula 
$n_\sigma \simeq N_{\rm excess}/\sqrt{N_{\rm bkg}}$, where $N_{\rm excess}$ is the 
number of excess events, and $N_{\rm bkg}$ is the  background estimate, whenever $N_{\rm excess} \ll {N_{\rm bkg}}$. For $N_{\rm excess} \leq {N_{\rm bkg}}$ the sensitivity calculation should be done in another premisses.
The significance of the Crab signal for one year is also shown in the bottom plot
of Fig.~~\ref{fig_sb2}.

%


\begin{figure}
\centering
\includegraphics[width=1.05\linewidth]{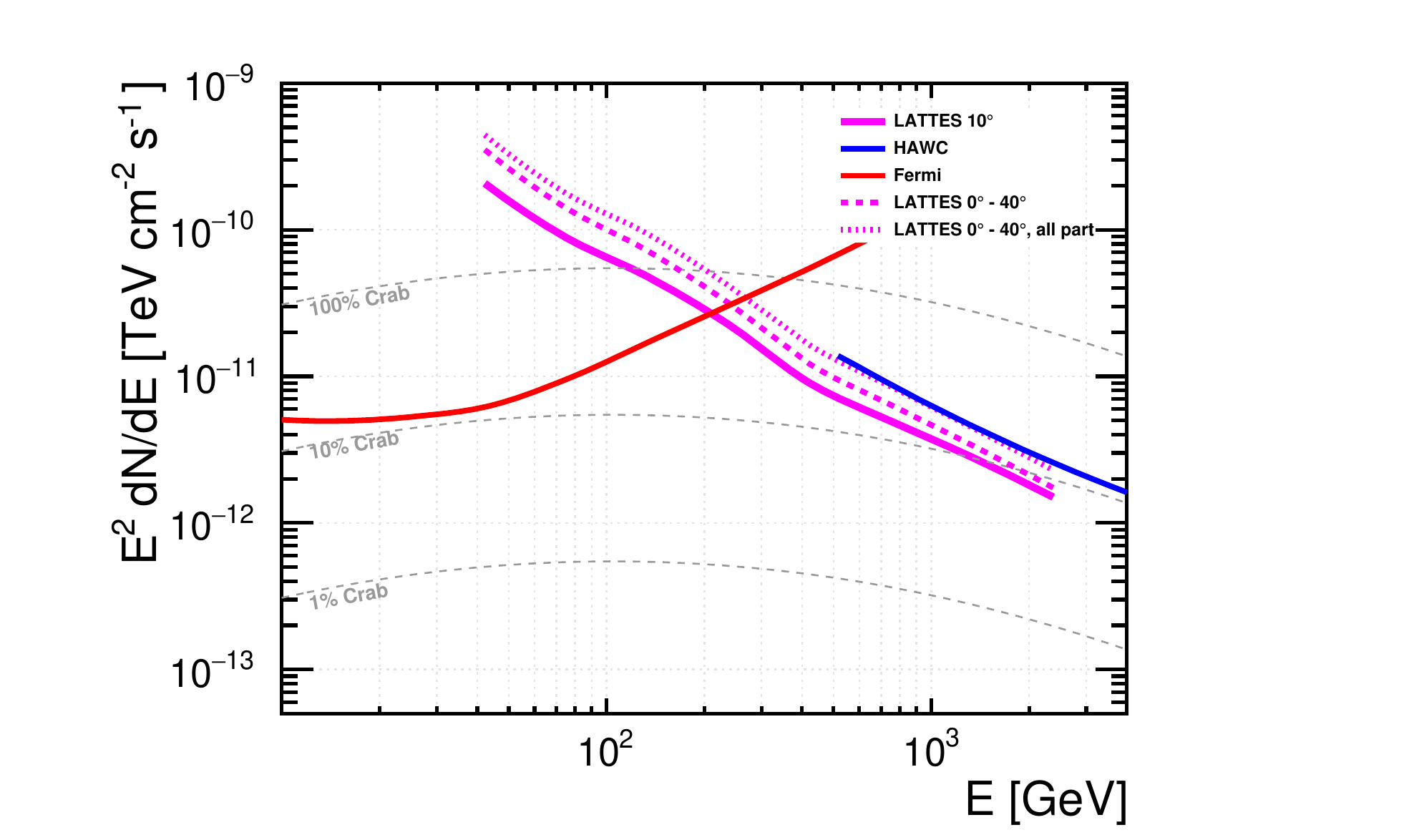}
\caption{
Differential sensitivity compared to the present sensitivity of  HAWC ~\cite{HAWC2017}  and the sensitivity of Fermi for observations of the galactic center~\cite{Fermi-10y}. 
The LATTES full line is the sensitivity obtained for a source at a zenith angle of 10$^\circ$; the dashed curve is an estimate of the LATTES performance while following the Crab Nebula transit; the thin dashed curve is an estimation of the sensitivity considering the all-particle cosmic-ray flux (see text for details).
For comparison, fractions of the Crab Nebula spectrum are plotted with 
the thin dashed gray lines.} 
\label{fig_diffsens}
\end{figure}

\subsection{Sensitivity for a steady source}\label{sec_diffsens}

To evaluate the performance of the detector, we compute  its differential 
sensitivity. To accomplish this we investigate the sensitivity in narrow bins of energy 
(4 bins per decade).
We compute the sensitivity  as the flux of a source giving 
$N_{\rm excess}/\sqrt{N_{\rm bkg}}=5$ after
1 year of effective observation time for a source visible for 1/4 of the time (this roughly 
corresponds to the visibility of the Galactic Centre  from the Southern tropic).

The result is shown in Fig. \ref{fig_diffsens}, and compared with the 
one-year sensitivities of $Fermi$ and HAWC~\cite{HAWC2017}. From this figure, it can be seen that this detector concept is in fact able to lower the energy threshold of present gamma-ray EAS experiments, with an acceptable sensitivity. It is worth noting at this point that this result comes from an end-to-end simulation.

While this result is solid, the comparison with  HAWC might not be fair as a source, like the Crab Nebula, is not  observed at a fixed zenith angle in the sky. In order to take this effect into account, one would need to simulate this detector concept for other zenith angles. Such task would not only be computationally heavy but it would also require the refinement of the shower reconstruction analyses to higher zenith angles, not to bias the performance results artificially. 

Hence, we decided to estimate the performance of this detector concept taking the full simulation at 10 degrees as baseline. 
CORSIKA photon shower simulations, equivalent to the ones generated at 10 degrees, were generated for zenith angles between $0$~degrees and 40~degrees. 
These CORSIKA showers were then processed by a fast detector simulation, where the amount of electromagnetic energy recorded by each station was saved. 
Under the assumption that the reduction of the sensitivity with the zenith angle arises mainly from the decrease of the shower trigger efficiency, due to the atmospheric attenuation,
a mapping between the effective area computed using the full simulation and the corresponding electromagnetic energy recorded at ground was created. 
With this mapping, the effective area of this array  can now be estimated  for other zenith angles.
It should be noted that this approach was undertaken only for the photon primaries. For the background, the effective area computed from the full simulation at 10~degrees was kept.
However, a reduction of the  flux of cosmic rays at ground, due to the increased atmospheric attenuation, is also expected. 
The effective area of the background at larger zenith angles is thus overestimated and the obtained sensitivity curve is probably a conservative one. 
Using the effective area as a function of the zenith angle and the time a source is at a given position in the sky one can easily compute the total sensitivity of the detector accounting for the transit of the source. 
To ease the comparison,  the Crab Nebula as seen from HAWC's latitude~\cite{astropy} was chosen. The obtained result can be seen in Fig.~\ref{fig_diffsens} as a dashed pink line. 
As expected  the degradation in sensitivity is more important at lower energies. At the highest energies the atmospheric attenuation barely affects the trigger efficiency and the only impact  
on the sensitivity comes from the reduction of the array area perpendicular to the source direction. 

Finally, it should be pointed that the simulated performance presented here uses for the background model the full simulation and reconstruction of the proton component. 
Although a full simulation of the heavier cosmic rays species was not considered here, an estimation of the expected sensitivity was also made using the all particle cosmic-ray flux  from~\cite{CRspectrum}. 
This is shown in Fig.~\ref{fig_diffsens} as  the thin dashed curve. It should be stressed that this is a conservative estimate. 
In fact, the number of muons and sub-showers increases with the mass of the primary cosmic-ray, allowing for a better discrimination of the hadronic background.
As such the  sensitivity, computed using a full simulation and reconstruction of the remaining background components,
is expected to be better than the curve shown above.

The differential sensitivity is independent of the spectral energy distribution (SED) of the emitting source. To compute the total sensitivity one must assume a SED; from this assumption, one can compute an integral sensitivity.

\begin{figure}
\centering
\includegraphics[width=1.05\linewidth]{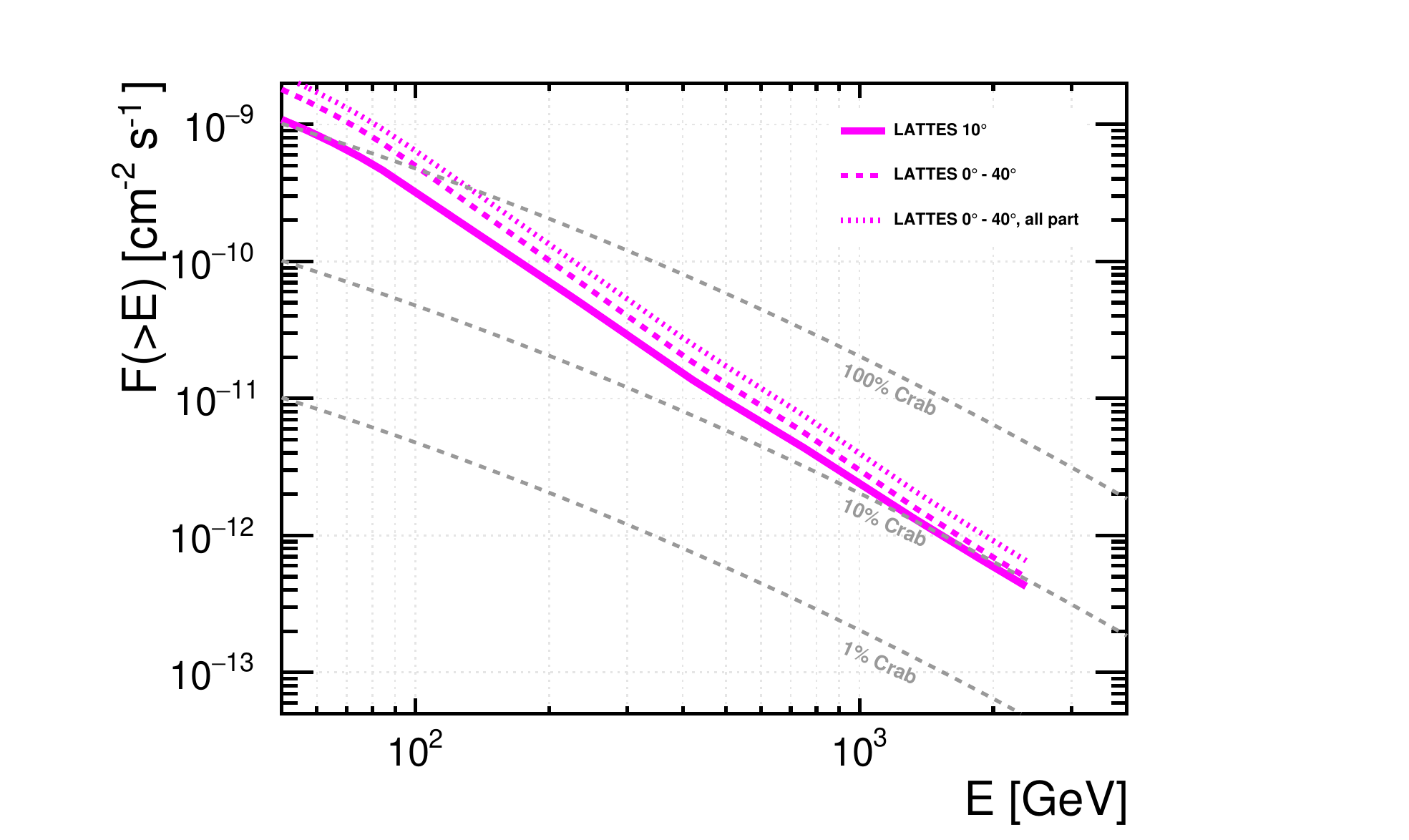}
\caption{
Integral sensitivity, defined as the flux of a source above a given energy for which $N_{\rm excess}/\sqrt{N_{\rm bkg}}=5$ after 1 year. The LATTES full line is the sensitivity obtained for a source at a zenith angle of $10^\circ$;
 the dashed curve is an estimation of the LATTES performance while following the Crab Nebula transit; the thin dashed curve is an estimation of the sensitivity considering the all-particle cosmic-ray flux (see text for details). For comparison, fractions of the integral Crab Nebula spectrum are plotted with the thin, dashed, gray lines.
}\label{fig_sens}
\end{figure}

We compute the integral sensitivity  as the flux of a source with a SED proportional to  the SED of the Crab Nebula giving $N_{\rm excess}/\sqrt{N_{\rm bkg}}=5$ after 1 year, and integrating all energies above a given energy. The integral sensitivity is shown in Fig.~\ref{fig_sens}.

\subsection{Discussion on the hybrid concept}

The hybrid nature of the proposed detector concept is the key ingredient for its performance, in particular at the lowest energies. 
It allows to decouple the calorimetric energy measurement of the shower particles at ground, using  the WCD detectors, from the
 measurement of their time, provided by the RPCs. 
In fact, the detector energy threshold can be lowered by increasing the light collection efficiency, and thus the total WCD signal.
This can be achieved by employing a highly reflective material to cover the inner walls of the WCDs. 
The main drawback is the increase of the light collection time and the consequent degradation of the WCD time resolution.
In our case, with the timing being provided by the RPCs, this is no longer an issue.

Furthermore, the  trigger is based solely on the compact array of small sized WCDs. 
Since RPCs are sensitive to  low energy charged particles, by removing the RPCs from the trigger, one avoids spurious triggers, namely 
due to the soil radioactivity. This allows also to reduce the energy threshold of the detector. 

\begin{figure}
\centering
\includegraphics[width=1.05\linewidth]{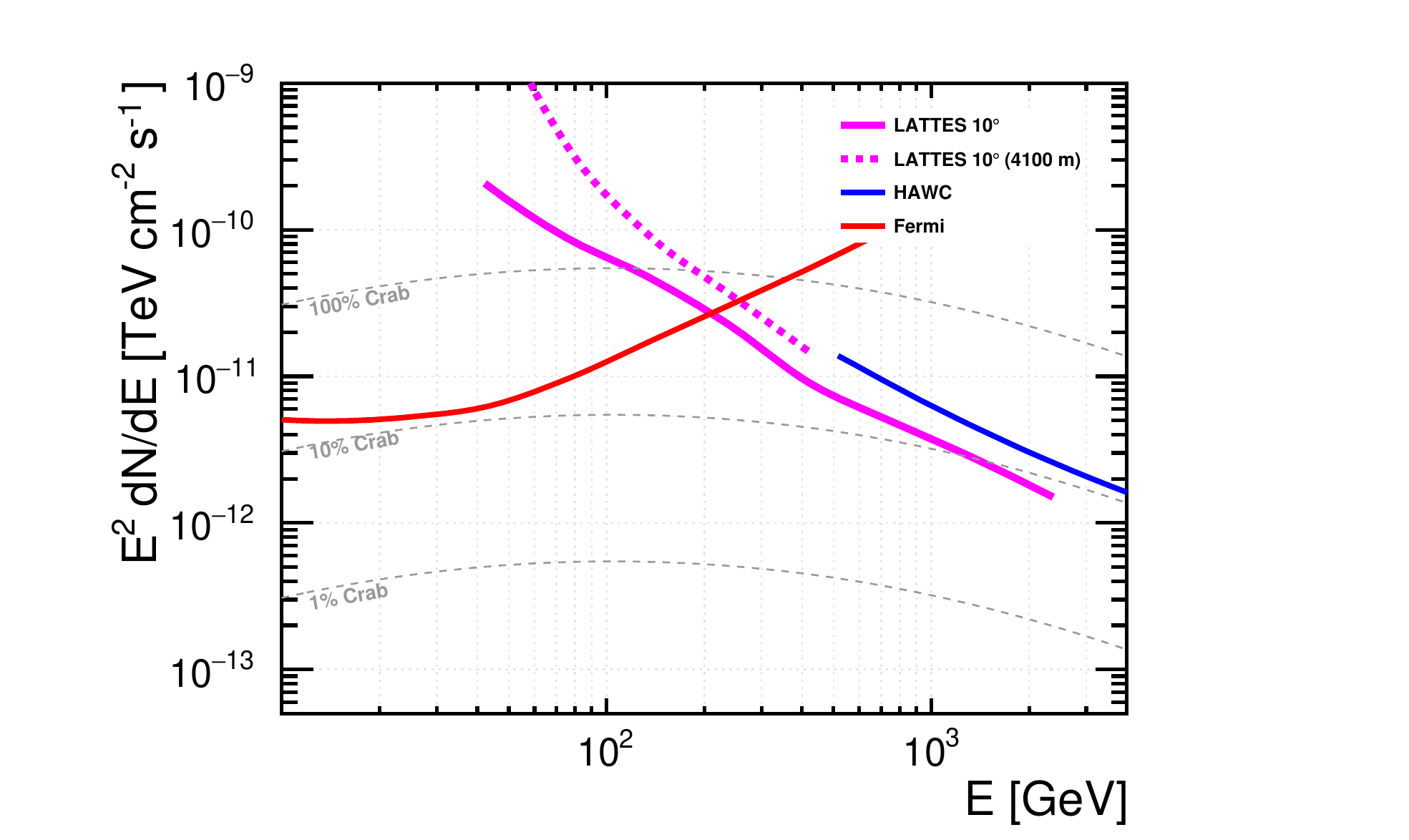}
\caption{
Same as figure~\ref{fig_diffsens}, including the LATTES sensitivity curve computed for an altitude of 4100 m.
}\label{fig_sens_4100}
\end{figure}

In order to better evaluate the impact of the present detector concept with respect to the gain expected  by going to high altitudes, 
the  differential sensitivity  of the array placed at 4100~m was computed from a full simulation for low energy showers and
is compared with the sensitivity at 5200~m in figure~\ref{fig_sens_4100}. 
Although a degradation of the performance is observed, as expected, the sensitivity is kept towards the low energies, down to 
about 100~GeV.

\section{Summary}\label{sec:concl}

We have proposed  a novel, large field-of-view, hybrid, extensive air shower detector 
sensitive to gamma rays of energies starting from 100 GeV (or even less for special high flux transients), based on individual 
units made (from top to bottom) by:
\begin{itemize}

\item a thin slab of lead, to allow conversion of secondary photons in showers into 
electron-positron pairs;

\item a position-sensitive detector like a RPC;

\item a water Cherenkov detector (or any other electromagnetic calorimeter with some 
muon identification capabilities).
\end{itemize}
 
We have shown that such a detector, if deployed on a surface of some 20~000 m$^2$, can reach a sensitivity that can provide the missing link between $Fermi$ and HAWC, between 100 and 350 GeV. Such detector would be able to detect in one year
with a $5\sigma$ significance a source as faint as the Crab Nebula at $100\,$GeV. Above $1\,$TeV it could reach sensitivities better than 10\% of the Crab Nebula flux.

The instrument is able to survey half of the sky, with enhanced ability to detect transient phenomena making it a very powerful tool to trigger observations of variable sources, in particular flares
by active galactic nuclei,  and to detect 
transients coupled to gravitational waves and gamma-ray bursts.


An external sparse array of units could be easily achieved, due to the modular nature of the detector. This would not only allow to extend 
the energy range but also would further improve the sensitivity of the core array at lower energies.

The presented hybrid detector concept is currently the baseline design of the core array of the LATTES project, foreseen as a new EAS gamma-ray detector in the Southern hemisphere, complementary to the planned Cherenkov Telescope Array.

\section*{Acknowledgements}
The authors  thank B. De Lotto for her comments on the manuscript.

R.~Concei\c{c}\~ao acknowledges the financial support given by Funda\c{c}\~ao para a Ci\^encia e Tecnologia, Portugal.  U. Barres de Almeida is funded by a CNPq Produtividade em Pesquisa 2 Grant Nr. 309306/2013-6 from Brazil. His work on this project is partially funded by a FAPERJ Thematic Grant Nrs. 210.951/2015 and 110.148/2013 from the State of Rio de Janeiro.



\bibliographystyle{woc}

\bibliography{lattes}

\end{document}